\documentclass[aps,prl, twocolumn,superscriptaddress,amsmath,amssymb]{revtex4}
\usepackage{graphicx}
\usepackage{graphicx}
\usepackage{textcomp}

\newcommand{\sto}{SrTiO$_{3}$}
\newcommand{\lao}{LaAlO$_{3}$}

\begin{document}

\title{Metal-Insulator Transition of the \lao-\sto\ Interface Electron System}

\author{Y.\ C.\ Liao}
\affiliation{Experimental Physics VI, Center for Electronic Correlations and Magnetism, Institute of Physics,
University of Augsburg, D-86135 Augsburg, Germany}
\author{T.\ Kopp}
\affiliation{Experimental Physics VI, Center for Electronic Correlations and Magnetism, Institute of Physics,
University of Augsburg, D-86135 Augsburg, Germany}
\author{C.\ Richter}
\affiliation{Experimental Physics VI, Center for Electronic Correlations and Magnetism, Institute of Physics,
University of Augsburg, D-86135 Augsburg, Germany}
\author{A.\ Rosch}
\affiliation{Institute of Theoretical Physics, University of Cologne, D-50937 Cologne, Germany}
\author{J.\ Mannhart}
\affiliation{Experimental Physics VI, Center for Electronic Correlations and Magnetism, Institute of Physics,
University of Augsburg, D-86135 Augsburg, Germany}


\date{\today}

\begin{abstract}
We report on a metal-insulator transition in the \lao-\sto\ interface electron system, of which the carrier
density is tuned by an electric gate field. Below a critical carrier density $n_\mathrm{c}$ ranging from
$0.5-1.5\times10^{13}/\mathrm{cm}^2$,  \lao-\sto\ interfaces, forming drain-source channels in field-effect devices are non-ohmic. The differential resistance at zero channel bias diverges within a 2$\%$ variation of the carrier density. Above $n_\mathrm{c}$, the conductivity of the ohmic channels has a metal-like temperature dependence, while below $n_\mathrm{c}$ conductivity sets in only above a threshold electric field. For a given thickness of the \lao\ layer, the conductivity follows a $\sigma_0  \propto (n -
n_\mathrm{c}) / n_\mathrm{c}$ characteristic. The metal-insulator transition is found to be distinct from that of the semiconductor 2D systems.
\end{abstract}

\maketitle

Conducting electron systems with unique properties can be generated at interfaces between highly insulating
oxides, the most widely studied case being the interface between the TiO$_2$-terminated (001) surface of \sto\
and \lao\ \cite{Ohtomo:2004}. This electron system behaves as a two-dimensional (2D) electron liquid
\cite{PhysRevB.81.153414} for which superconducting \cite{Reyren:2007} and magnetic ground states
\cite{Brinkmann:2007} have been reported. At low temperatures, the interface can be tuned from a superconducting
to a resistive state by applying transverse electric fields \cite{Caviglia2008}. At higher temperatures, large
electric gate fields drive the system through a metal-insulator transition (MIT) \cite{thiel2006,cen2008}.

The \lao-\sto\  2D system is disordered \cite{PhysRevLett.102.046809}; the disorder arising, for example, from dislocations crossing the interface, or from point defects. Consequently, as a 2D electron system (2DES), the
interface is expected to be an insulator for $T=0$, at least for negligible interaction strength among the electrons
\cite{PhysRevLett.42.673}. The observed metallic behavior may be interaction-driven or a crossover effect; sizable
interactions  have indeed to be anticipated due to the large interaction energy at small carrier
densities~\cite{Lu10,commentXC}. Indeed, for  the 2DES of semiconductor
interfaces~\cite{PhysRevB.51.7038,Kravchenko2004} it has been argued~\cite{PhysRevLett.79.455} that the metallic phase is not a Fermi liquid because a fictitious electron system with suppressed electronic
correlations would form a localized phase \cite{PhysRevLett.42.673} rather than a 2D Fermi gas.
The polar nature of the \lao-\sto\ interface possibly results in additional defects and
excitations. Excitations of localized electrons and charged defects, for example, can enhance dephasing and thereby reduce weak localization.

The  MIT at semiconductor interfaces is still being debated intensely
\cite{PhysRevLett.79.1543,PhysRevLett.83.3506,PhysRevLett.83.4642,AltshulerMaslov99,Washburn1999,PhysRevB.79.235307}. The discovery of a MIT at the
interface of perovskite oxides, an entirely different host structure for a 2DES, may shed light on the nature of
the MIT in two dimensions.  In previous experiments \cite{thiel2006,cen2008} the MIT was deduced from the
suppression of the conductance, but has not been investigated further.

In field-effect studies of the perovskite interfaces, the gate fields were considered to change the properties of the electron system primarily by altering the carrier density $n$. It has recently been revealed, that by compressing the electron wave function toward the
interface, the field also changes the effective disorder of the system and therefore its electronic mobility
$\mu$ \cite{PhysRevLett.103.226802}.

To advance these issues further, we have measured the current-density vs.\ electric-field characteristics ($J(E)$) of the interface electron system as a function of applied gate fields. These studies show that in the samples investigated the effects of the gate fields have a strong component that arises from a change of $n$. The data furthermore reveal a power-law behavior of the conductivity $\sigma$ of the 2D electron liquid as function of $n$.

\begin{figure}[b]
\includegraphics[width=.6\columnwidth,clip]{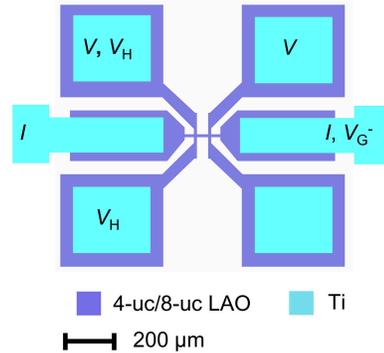}
\vspace{0mm} \caption[]{\label{device} Sketch of the sample structure (top view). }
\end{figure}

For the measurements we have fabricated seven samples with thicknesses of the \lao\ layers of 4 and 8 unit cells
(uc) to investigate a possible dependence of the field effects on the \lao\ thickness. The thickness of 4 uc was
chosen to yield the critical thickness required for interface conduction \cite{thiel2006}. The samples were
fabricated by epitaxially growing \lao\ layers on the (001) surfaces of TiO$_2$-terminated
\cite{Kawasaki12021994,koster:1998}, 1 mm thick \sto\ single crystals. Scanning force microscopy was used to
verify that the substrate surfaces were atomically flat. The films with a nominal composition of \lao\ were
grown by pulsed laser deposition from a single crystalline \lao\ target using standard deposition conditions
(780~$^{\circ}$C, $7\times10^{-5}$ mbar O$_2$). By reflective high energy electron diffraction the thickness of
the \lao\ layers was controlled with a precision of $\sim$ $0.1 - 0.2$ uc.  After growth, the films were
annealed for an hour in 400 mbar O$_2$ to minimize oxygen vacancies. The devices were photolithographically
patterned \cite{schneider2006} into the structure shown in Fig.~\ref{device}. This structure was optimized for
Hall measurements and for four-point measurements of the $J(E)$-characteristics, using a small distance between
the voltage contacts to minimize the inhomogeneity of the gate fields. Contact pads for the voltage and current
leads were prepared by sputtering Ti into holes Ar-ion etched through the \lao. The back gate was provided by
silver diffused into the \sto.

For each gate voltage ($V_\mathrm{G}$) the $J(E)$-characteristic in zero magnetic field, Hall voltages and the
magnetoconductances at several gate voltages were measured. Gate leakage was always below 1~nA. Only negative
gate voltages were applied, for these characteristics were found to be reversible with $n$.

\begin{figure}[t]
\includegraphics[width=.9\columnwidth,clip]{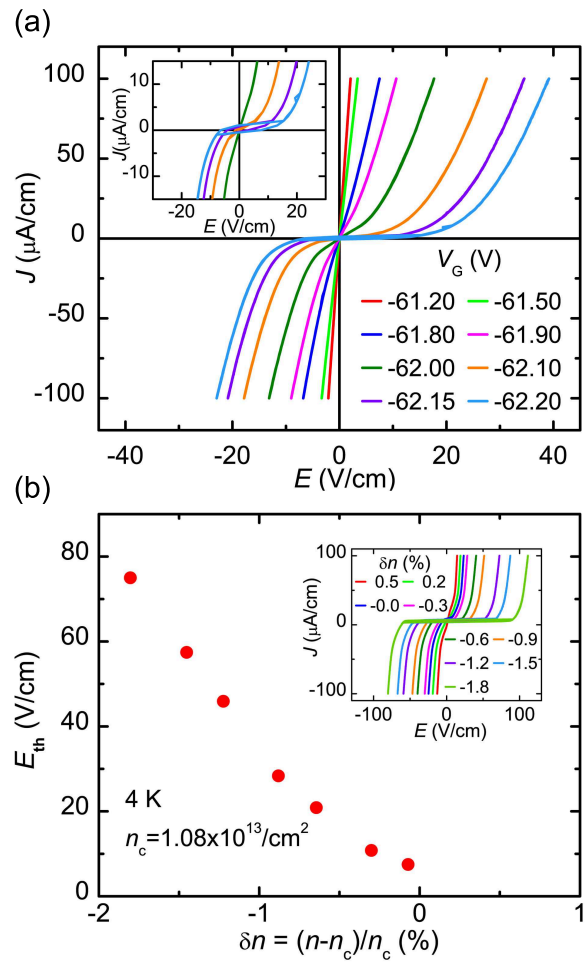}
\vspace{0mm} \caption[]{\label{IV} (a) $J(E)$-characteristics of a measurement bridge in a \lao-\sto\ interface
with a 4~uc thick \lao\ layer (4~K, no applied magnetic field). The bridge was 60~$\mu$m long and 10~$\mu$m wide. At $V_\mathrm{G}=-\,61.2$~V, the $J(E)$-characteristic changes from linear to non-linear behavior with a much smaller conductivity. At $V_\mathrm{G}=-\,62.1$~V, the differential conductivity at $J\sim 0$ equals
$\sim10^{-7}$~S. Below $V_\mathrm{G}=-\,62.1$~V a hysteresis emerges near zero bias (inset) and conductivity sets in only above a threshold field $E_{\rm th}$. Because in the
insulating regime the voltage $V$ along the bridge is no longer small compared to $V_\mathrm{G}$, the respective $J(E)$-characteristics are asymmetric.  (b) Threshold electric field $E_{\rm th}$
plotted as function of the reduced carrier density $\delta n=(n - n_\mathrm{c}) / n_\mathrm{c}$. The inset shows the respective $J(E)$ curves.}
\end{figure}

The $J(E)$-characteristics are shown in Fig.~\ref{IV}(a). At small $J$ and small $|V_\mathrm{G}|$
($V_\mathrm{G}\geq -\,61.2$~V, 4~K) the characteristics are linear for the whole range of $J$ and for \mbox{$T<
50$}~K. At larger $|V_\mathrm{G}|$ (\mbox{$V_\mathrm{G}<-\,61.2$}~V, 4~K), however, the $J(E)$-characteristics
are nonlinear, showing an enhanced differential resistance for small currents. For
\mbox{$V_\mathrm{G}<-\,62.1$}~V (inset of Fig.~\ref{IV}(a)), the curves show a clear threshold behavior:
below a characteristic threshold field  $E_\mathrm{th}$,
the current density is extremely small ($<2$~$\mu$A/cm)
and shows a weak hysteretic behavior, which is caused by the $RC$-time constant of the measurement.  Because these current densities are minute, we cannot rule out
that they are affected by finite gate currents.
Above  $E_\mathrm{th}$ the current grows nonlinearly. The width $E_\mathrm{th}$  diminishes with increasing $n$ to approach zero at a critical density $\sim n_\mathrm{c}$ (Fig.~\ref{IV}(b)) defined below.

\begin{figure}[t]
\includegraphics[width=.9\columnwidth,clip]{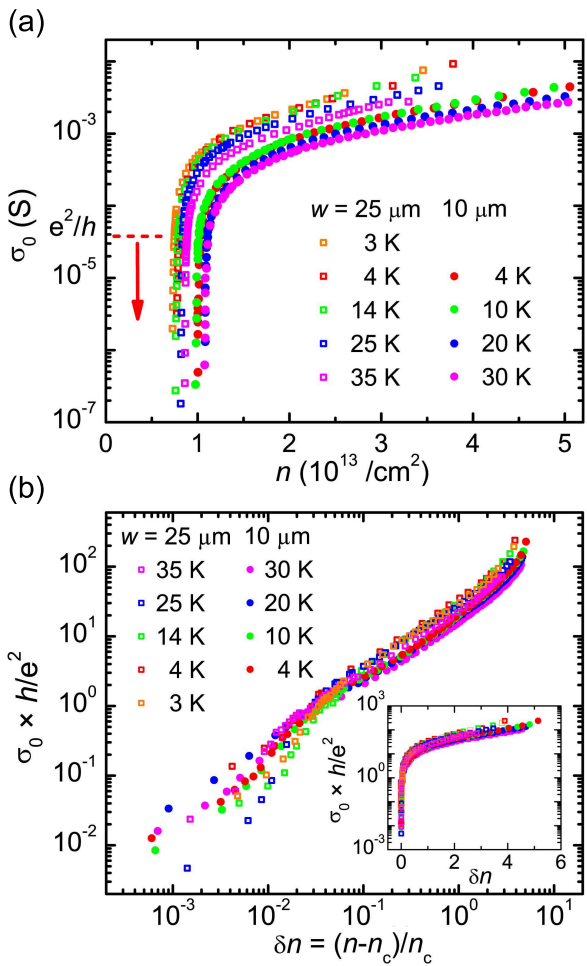}
\vspace{0mm} \caption[]{\label{Sigmascaling} (a) Conductivities of 10~$\mu$m and 25~$\mu$m wide channels in
4 uc \lao-\sto\ interfaces measured as function of the carrier density varied by $V_\mathrm{G}$  (in zero
magnetic field). For $\sigma_0\lesssim e^2 /h$, the current-voltage characteristics are non-linear. Only data points derived from $J(E)$
curves that are reversible for the whole bias range are plotted. (b) Conductivities of \lao-\sto\ interfaces
plotted as function of the reduced carrier density. The scaling curve only includes data points in the ohmic
regime. For $\delta n=0$, $\sigma_0$ becomes minute (inset).}
\end{figure}

The conductivity curves bear the characteristic shape shown in Fig.~\ref{Sigmascaling}(a) for the 4 uc thick
samples in zero magnetic field. The 8 uc thick samples display a similar, but noisier behavior (shown in
supplement). For $n\geq 2\times10^{13}/\mathrm{cm}^2$ the conductivity at zero bias $\sigma_0$ decreases
approximately linearly with $n$. As long as the samples show linear~\cite{comment0} $J(E)$-characteristics, e.g.\
at \mbox{$V_\mathrm{G}>-\,61.2$}~V (Fig.~\ref{IV}(a)), their conductivities are at least of order $e^2 /h$.
Conductivity values with $\sigma_0\lesssim e^2 /h$ that are plotted in Fig.~\ref{Sigmascaling}(a) were obtained
from non-linear characteristics and present the differential conductivity at zero bias ($\sigma_0$). For
$n<n_\mathrm{c}\sim 0.5-1.5\times10^{13}/\mathrm{cm}^2$ (depending on sample and temperature), however,  $\sigma_0$ collapses and the samples are
effectively insulating (Fig.~\ref{IV}), unless an in-plane electric field $E>E_\mathrm{th}$ is applied. The
transition from the insulating to the linear regime occurs within $\sim 0.02\, n_\mathrm{c}$ and is reversible
with $n$. The Hall mobility $\mu\equiv\sigma/n e$ is small near the transition, equalling $\sim 30\,
\mathrm{cm}^2/\mathrm{Vs}$ and $\sim 5\, \mathrm{cm}^2/\mathrm{Vs}$ as calculated in the linear regime for the 4
and 8~uc thick samples, while outside the transition $\mu$ reaches 1000~cm$^2$/Vs.

Overall, the temperature dependence of the conductivity is surprisingly weak and can, to a large extent,
be absorbed into a simple shift of the curves along the $n$-axis.
For all samples of a given thickness we therefore find that the $\sigma (n)$-characteristics in the ohmic regime can be
scaled onto a master curve (Fig.~\ref{Sigmascaling}(b)). In this figure the conductivity $\sigma_0$ is plotted as
a function of the reduced carrier density $\delta n=(n - n_\mathrm{c}) / n_\mathrm{c}$. The values of
$n_\mathrm{c}$ depend on temperature and are given in the supplementary materials section.
For $0.05<\delta n<1$, in this range $\sigma_0 > e^2/h$, the master curve is
characterized by a power-law behavior $\sigma_0\propto\delta n^q$, with $q \sim 1$. Interestingly, the interfaces follow this characteristic curve for the whole
temperature range for which the experiments could be performed (4.2~K~$ < T < 50$~K), although the dielectric
constant of \sto\ is temperature dependent. While the data of the 8~uc samples are characterized by considerably
larger scatter, for $0.05<\delta n<1$ they approach the curve of the 4 uc thick samples (shown in the
supplement).

In the vicinity of the MIT, within a  2$\%$ variation of the carrier density, the data scatter more and the zero-bias differential conductivity deviates from a power-law dependence, when $\sigma_0 < e^2/h$. Similarly, on the insulating side a decrease of $n$ by just $1$ or $2$\% causes substantial threshold fields (Fig.~\ref{IV}b).

For an analysis of the behavior close to the MIT, it is suggestive to compare the 2DES of \lao-\sto\ interfaces
to the 2DES of semiconductor interfaces \cite{PhysRevB.51.7038,Kravchenko2004,PhysRevLett.79.1543,Washburn1999}, for which many different scenarios have been discussed, see, e.g., \cite{PhysRevLett.79.455,PhysRevLett.79.1543,Washburn1999,PhysRevLett.83.3506,AltshulerMaslov99,PhysRevLett.83.4642}.
In these semiconductor systems, the transition
is controlled by {\em two} parameters, temperature and density, whose interplay can be described
by a single-parameter scaling function, $\sigma (\delta n,T)=f(\delta n/T^{1/\nu z})$, where $\nu$ is the correlation
length exponent, and  $z$ the dynamical critical exponent. These exponents are usually of the order of one.  In contrast, the conductivity of the \lao-\sto\ interfaces
is primarily controlled by $\delta n$; the temperature has very little influence and accordingly such a
scaling relation is not found.  In this
respect, the MIT is not comparable to that of the semiconductor interfaces. Moreover, the $J(E)$-characteristics for $n<n_\mathrm{c}$ display a threshold behavior with a subsequent  $J(E)\propto (E-E_\mathrm{th})^2$ dependence which has not been observed for the semiconductor interfaces.

Several candidates of possible insulating states exist. The first candidate is an Anderson insulator, in which single electrons are localized by disorder. The second candidate arises, if Coulomb interactions are strong; the electrons then form a Wigner crystal which is pinned by disorder. The third candidate is a polaronic insulator; strong electron-phonon interaction can lead to a self-trapping and localization. One of our surprising results is the absence of significant temperature dependencies of the $J(E)$-characteristics in combination with strong non-linear effects from moderate electric fields. In a typical electric field of 10\,V/cm, an electron has, for example, to travel about 1\,$\mu$m, more than 100 times the typical distance of electrons at $n_c$, to gain an energy of $k_B T$ (for $T=10$~K). The effective absence of thermal effects
therefore implies that either barriers prohibiting transport are macroscopic in size or that the motion of electrons is controlled by a collective effect in which many electrons participate as is, e.g., the case in a Wigner crystal. Indeed, the characteristic dependence of the conductivity on the electric field
is known from the depinning of two-dimensional vortex lattices in superconducting films
\cite{Kes1983}. Also simple estimates \cite{commentXC} suggest that Coulomb interactions can be
sufficiently strong to induce Wigner crystallization, especially if one takes into account that polaronic effects
may enhance the tendency to crystallization.

A possible scenario is therefore that at the MIT an insulating phase, such as a pinned Wigner crystal,
percolates through the system. Remarkably, the MIT has been related to a percolation transition also in semiconductor systems
\cite{PhysRevB.79.235307,PhysRevLett.83.3506}. The critical conductance exponent of a percolation transition  in
2D is 1.3, in conformity with our findings. For the low electronic densities
in the semiconductor systems, the transition is possibly induced through scattering by unscreened impurity
charges \cite{PhysRevB.79.235307}. For the \lao-\sto\  2D system the charge density of
order $10^{13}/\mathrm{cm}^2$ is so high that it may preclude the observation of  a percolation transition
driven by unscreened impurity charges.  An alternative possibility is a transition
where a Wigner crystal or another phase, e.g., a glassy state induced by strong Coulomb interactions
\cite{PhysRevLett.83.4642}, melts. It is also noted that for oxide materials $n_c$ is
a small carrier density. The carrier injection induced by a electric field $E>E_{\rm th}$ causes a
non-equilibrium state with a carrier density that is locally enhanced. We therefore regard it as a possibility
that the MIT is influenced or even triggered by macroscopic inhomogeneities in the sample and
the enhanced carrier density at the contacts. This could also affects the $J(E)$-dependencies,
causing non-linearities.

\begin{figure}[t]
\includegraphics[width=.8\columnwidth,clip]{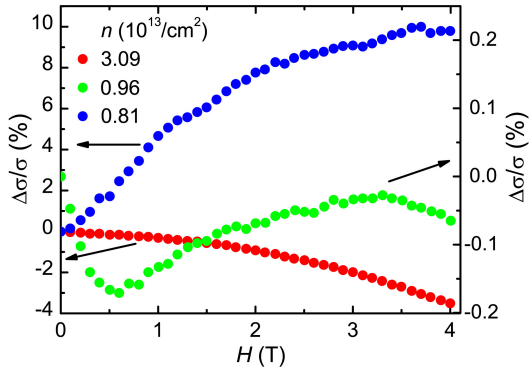}
\vspace{0mm} \caption[]{\label{Magneto} Magnetoconductance of the 25~$\mu$m sample shown in
Fig.~\ref{Sigmascaling}(a) measured at 4.2~K. Due to the large Hall voltage in the depleted state, the
conductance is symmetrized using $(\sigma (H) + \sigma (-H))/2$. From high to low carrier density, the
corresponding reduced carrier densities, $\delta n$, are 2.96, 0.23, and 0.03, respectively.} \end{figure}

To illustrate the characteristics of the \lao-\sto\ interface above the MIT,  typical magnetoconductance curves
are shown in Fig.~\ref{Magneto}. At high carrier density, the $\Delta \sigma /\sigma$-characteristics exhibit
a $-H^2$ dependence (red curve), resembling the typical behavior of metals in small magnetic fields. As revealed
in previous studies \cite{bell:222111,Caviglia2008,PhysRevLett.104.126802,PhysRevLett.104.126803}, the
magnetoconductance of the depleted \lao-\sto\ interface shows a negative derivative in low fields and a positive in high fields (green curve). This behavior has been interpreted as the result of spin-orbit interaction and weak localization  \cite{PhysRevLett.104.126803}. Near the MIT, in contrast,
the magnetoconductance is always positive and has a negative second derivative in high fields (blue curve).

In summary, the transport properties of \lao-\sto\ interfaces were measured as a function of the mobile carrier
density altered by gate fields. It is found that below a critical carrier density
$n_\mathrm{c}\sim1\times10^{13}/\mathrm{cm}^2$ the interfaces become insulating for small in-plane electric
field. For samples with 4~uc thick \lao\ layers, this transition occurs at a conductivity of order $e^2 /h$. For
a wide range of temperatures, the dependencies of the conductivity on the reduced carrier density
follow a $\sigma_0 \propto(n - n_\mathrm{c}) / n_\mathrm{c}$ characteristic.

While it remains to be explored, which of the proposed microscopic mechanisms, such as disorder potentials or polaronic enhancement of Wigner localization, determine the nature of the MIT, it is evident that
the MIT  in \lao-\sto\ interfaces differs qualitatively from the one in two-dimensional semiconductor systems.

\begin{acknowledgments}
The authors gratefully acknowledge fruitful discussions with T. Nattermann and  financial support by the DFG (TRR 80, SFB 608) and the EU (oxIDes).
\end{acknowledgments}


\begin{thebibliography}{30}
\bibitem[{com({\natexlab{a}})}]{commentXC}
\bibinfo{note}{At the \lao-\sto\ interface, the Fermi energy equals
  approximately 10~meV~\protect\cite{comment2}. Below 50~K, the oxide
  interfaces therefore do not reach the Fermi temperature. The kinetic energy
  per particle is $E_\mathrm{kin} \propto n$ and the Coulomb interaction $V\propto
  n^{1/2}$, so that $V/ E_\mathrm{kin} \sim 100/\varepsilon$. With a dielectric
  constant $\varepsilon\sim 5-10$ we conclude that the Coulomb energy exceeds
  the kinetic energy by an order of magnitude, similar to the semiconductor
  systems.}
\bibitem[{com({\natexlab{b}})}]{comment0}
\bibinfo{note}{We use the word ``ohmic'' following the usage to describe $J(E)$
  characteristics with a finite conductance at $E=0$. $J(E)$ characteristics in
  which $J\propto E$ are called ``linear''.}
\end{thebibliography}


\end{document}